\documentclass[aps,prd,twocolumn,10pt,groupedaddress]{revtex4-1}
\usepackage{amssymb}
\usepackage{graphicx}
\usepackage{amsmath}
\usepackage{hyperref}

\begin{document}

\title{Dark matter superfluid and DBI dark energy}
\author{Rong-Gen Cai}
\email{cairg@itp.ac.cn}
\author{Shao-Jiang Wang}
\email{schwang@itp.ac.cn}
\affiliation{State Key Laboratory of Theoretical Physics, Institute of Theoretical Physics, Chinese Academy of Sciences, Beijing 100190, China}
\date{\today}

\begin{abstract}
  It was shown recently that, without jeopardizing the success of the $\Lambda$ cold dark matter model on cosmic scales, the modified Newtonian dynamics (MOND) can be derived as an emergent phenomenon when axionlike dark matter particles condense into superfluid on the galactic scales. We propose in this paper a Dirac-Born-Infeld (DBI) scalar field conformally coupled to the matter components. To maintain the success of MOND phenomenon of dark matter superfluid on the galactic scales, the fifth force introduced by the DBI scalar should be screened on the galactic scales. It turns out that the screening effect naturally leads to a simple explanation for a longstanding puzzle that the MOND critical acceleration coincides with present Hubble scale. This galactic coincidence problem is solved, provided that the screened DBI scalar also plays the role of dark energy on the cosmic scales.
\end{abstract}
\maketitle

\section{Introduction}

Recently, a novel theory of dark matter (DM) superfluidity \cite{Berezhiani:2015pia,Berezhiani:2015bqa} was proposed to combine the success of modified Newtonian dynamics (MOND) \cite{Milgrom:1983ca,Milgrom:1983pn,Milgrom:1983zz} on galactic scales with the triumph of the $\Lambda$ cold dark matter ($\Lambda$CDM) on cosmic scales. The MOND turns out to be an emergent phenomenon of DM itself on galactic scales due to a MOND-like force between baryons mediated by superfluid phonons of the axionlike particles condensed as superfluid with a coherence length of order the galactic size and a critical temperature of order micro-Kelvin. The $\Lambda$CDM model is eventually recovered beyond galactic scales when the fraction of particles in the condensate decreases with increasing temperature due to larger velocity dispersion and hence larger DM temperature in galaxy clusters.

It was known as the galactic coincidence \cite{Famaey:2011kh} that a critical acceleration scale appears in various seemingly unrelated Kepler-like laws of galactic dynamics, which cannot be simply explained in a common way in the context of the cold dark matter (CDM) scenario. However, MOND predicts such a universal acceleration scale $a_0\approx10^{-10} \mathrm{m/s^2}$, which should intriguingly happen to be of order the present Hubble scale $H_0\sim a_0$ or more boldly the cosmological constant scale $\Lambda^4\sim M_{\mathrm{Pl}}^2a_0^2$. Although MOND now  emerges  from DM itself on galactic scales in the context of DM superfluidity, the galactic coincidence still manifests itself as an input parameter in order to fix other parameters to their preferred values. It should be in any case striking that the dark matter and dark energy sectors have such a common scale even though it is currently unclear whether it is just a coincidence or smoking gun for new physics.

It was also known as the cosmic coincidence that the energy density used to account for the late-time cosmic acceleration happens to be the same order of magnitude as the matter components today. Alternative to the standard cosmological constant scenario, one might as well consider a slowly rolling scalar field known as dynamical dark energy (DE) with proper screening mechanisms \cite{Joyce:2014kja} to hide the fifth force from the local tests of gravity. To at least alleviate the cosmic coincidence, the energy density in the scalar field should at least track \cite{Zlatev:1998tr,Steinhardt:1999nw} the background energy density and then grow to dominate the energy budget at late times. Either the screening mechanism or tracking behavior can be realized if general interactions between dark energy and matter components are concerned.

In this paper, we propose a very simple explanation for the galactic coincidence problem by conformally coupling a Dirac-Born-Infeld (DBI) scalar field with local matter components. To effectively screen the fifth force mediated by the DBI scalar field from the MONDian force mediated by DM superfluid phonons on galactic scales, the galactic coincidence $a_0=\Lambda^2/2gM_{\mathrm{Pl}}\sim H_0$ is derived, provided that the DBI characteristic scale $\Lambda^4\sim M_{\mathrm{Pl}}^2H_0^2\sim(\mathrm{meV})^4$ coincides with current critical energy density for conformal coupling $g\sim\mathcal{O}(1)$. This allows us to interpret the DBI scalar field as a dynamical DE in the presence of a conformal coupling term. The equation of state (EOS) of our DBI dark energy mimics that of Chaplygin gas.

This paper is organized as follows. In Sec. \ref{sec:2}, we review the DM superfluidity and define the MOND transition scale. In Sec. \ref{sec:3}, we propose a DBI-like scalar conformally coupled with the matter component to solve the galactic coincidence problem. In Sec. \ref{sec:4}, the possibility of our DBI scalar playing the role of DE is explored. The final section is devoted to conclusions and discussions.

\section{Dark matter superfluid}\label{sec:2}

In the nonrelativistic regime, DM superfluid \cite{Berezhiani:2015pia,Berezhiani:2015bqa} is effectively described by the MOND Lagrangian with a conformal coupling term to baryons,
\begin{equation}\label{eq:MONDTb}
\mathcal{L}_{\mathrm{MOND}T_\mathrm{b}}=\frac{2}{3}\Lambda(2m)^{3/2}X\sqrt{|X|}+\frac{\alpha\Lambda\theta}{M_{\mathrm{Pl}}}T_{\mathrm{b}},
\end{equation}
where DM particle $m$ is of order $\mathrm{eV}$ to ensure the formation of Bose-Einstein condensation and the phonon excitation $X=\dot{\theta}-m\Phi-(\vec{\nabla}\theta)^2/2m$ is described by the Goldstone boson $\theta$ for a spontaneously broken global $U(1)$ symmetry under the external gravitational potential $\Phi$. The dimensionless parameter $\alpha$ and dimensionful parameter $\Lambda$ can be fixed later by inputting the MOND critical acceleration $a_0$ in order to reproduce the MONDian profile. For static spherically symmetric profile $\theta=\mu t+\varphi(r)$ at constant chemical potential $\mu$ and baryons distribution $T_{\mathrm{b}}=-\rho_{\mathrm{b}}(r)$, the equation of motion (EOM)
\begin{equation}
\frac{1}{r^2}\frac{\partial}{\partial r}\left(r^2\sqrt{2m|X|}\varphi'(r)\right)=\frac{\alpha\rho_{\mathrm{b}}(r)}{2M_{\mathrm{Pl}}}
\end{equation}
can be integrated for the $X<0$ branch to obtain
\begin{equation}
\varphi'(r)\simeq\sqrt{\frac{\alpha M_{\mathrm{b}}(r)}{8\pi M_{\mathrm{Pl}}r^2}}\equiv\sqrt{\kappa}
\end{equation}
for $\kappa\gg\mu-m\Phi$ with $M_{\mathrm{b}}(r)\equiv4\pi\int_0^rr'^2\mathrm{d}r'\rho_{\mathrm{b}}(r')$, which admits a MONDian acceleration,
\begin{equation}
a_{\varphi}=\alpha\frac{\Lambda}{M_{\mathrm{Pl}}}\varphi'\simeq\sqrt{\frac{\alpha^3\Lambda^2}{M_{\mathrm{Pl}}}\frac{GM_{\mathrm{b}}(r)}{r^2}},
\end{equation}
if one identifies
\begin{equation}
\frac{\alpha^3\Lambda^2}{M_{\mathrm{Pl}}}\equiv a_0,
\end{equation}
hence $\alpha\sim\mathcal{O}(1)$ for $\Lambda\sim\mathrm{meV}$.

The general picture of DM superfluidity is that the DM halo core where galaxies are located is almost entirely condensed and the dynamics is dominated by the MONDian force mediated by the DM superfluid phonons, whereas galaxy clusters are either in a mixed phase or entirely in the normal phase just as those on cosmic scales. Therefore, it is natural to define a MONDian transition radius
\begin{equation}\label{eq:MONDr}
r_{\mathrm{MOND}}=\sqrt{\frac{MG}{a_0}}
\end{equation}
in the context of the DM superfluid core with core radius $r_{\mathrm{MOND}}$ containing the total mass of $M$. To see that this is a reasonable definition, consider a DM halo with central density $\rho_0\sim M_{r_0}/r_0^3$ and core radius $r_0=\sqrt{M_{r_0}G/a_0}$; one obtains a constant surface density $\rho_0r_0\sim M_{r_0}/r_0^2\sim a_0/G$ independent of galaxy luminosity found recently by several astrophysical observations \cite{Kormendy:2004se,Spano:2007nt,Donato:2009ab,Gentile:2009bw}. One can even reproduce a sort of baryonic Tully-Fisher relation (BTFR) \cite{McGaugh:2000sr,McGaugh:2005qe,McGaugh:2011ac} $M_{r_0}\sim\rho_0r_0^3\sim(a_0/G)r_0^2\sim v^4/Ga_0$ by using $\rho_0r_0\sim a_0/G$ and $a_0\sim v^2/r_0$. The MONDian transition radius thus serves as a natural separation between the MOND regime $r<r_0$ with $a_{\mathrm{N}}<a_0$ and the Newtonian regime $r>r_0$ with $a_{\mathrm{N}}>a_0$ where $a_{\mathrm{N}}=GM_r/r^2$.

\section{DBIonic screening}\label{sec:3}

The action of the scalar field we propose in this paper has the form
\begin{align}\label{eq:DBITm}
\nonumber S_{\mathrm{DBI}T_{\mathrm{m}}}=&\int\mathrm{d}^4x \sqrt{-f}\left(-\Lambda^4\sqrt{1-\Lambda^{-4}(\partial\phi)^2}\right)\\
&+\int\mathrm{d}^4x \sqrt{-f}\frac{g\phi}{M_{\mathrm{Pl}}}T_{\mathrm{m}},
\end{align}
which will be referred to as the $\mathrm{DBI}T_{\mathrm{m}}$ action for short. It should be kept in mind that the same symbol $\Lambda$ used in our action (\ref{eq:DBITm}) has nothing to do with that in the action (\ref{eq:MONDTb}), although they actually coincide as we will see later. Here, $f$ is the determinant of the Friedmann-Robertson-Walker (FRW) metric of a 3-brane moving in a five-dimensional Minkowski space with two time dimensions,
\begin{equation}
\mathrm{d}s_5^2=-\mathrm{d}w^2+f_{\mu\nu}\mathrm{d}x^{\mu}\mathrm{d}x^{\nu}.
\end{equation}
Here, the Gaussian normal transverse coordinate $w(x)=\Lambda^{-2}\phi(x)$ is written in terms of the DBI scalar field $\phi(x)$. The first term in $\mathrm{DBI}T_{\mathrm{m}}$ action (\ref{eq:DBITm}) can thus be interpreted as a cosmological constant term,
\begin{equation}
S=\int\mathrm{d}^4x\sqrt{-g}(-\Lambda^4)=\int\mathrm{d}^4x\sqrt{-f}(-\Lambda^4\gamma^{-1}),
\end{equation}
in terms of the induced metric $g_{\mu\nu}=f_{\mu\nu}-\Lambda^{-4}\partial_{\mu}\phi\partial_{\nu}\phi$ on the brane, and the inverse of the induced metric is just $g^{\mu\nu}=f^{\mu\nu}+\Lambda^{-4}\gamma^2\partial^{\mu}\phi\partial^{\nu}\phi$ with an abbreviation $\gamma\equiv1/\sqrt{1-\Lambda^{-4}(\partial\phi)^2}$.

The first term in (\ref{eq:DBITm}) differs from the standard DBI action \begin{equation}
S_{\mathrm{DBI}}=\int\mathrm{d}^4x \sqrt{-f}\left(-\Lambda^4\sqrt{1+\Lambda^{-4}(\partial\phi)^2}\right)
\end{equation}
by a flipped sign in front of the derivative term, which as we will see is essential for the so-called DBIonic screening mechanism \cite{Burrage:2014uwa}. It is worth noting that the first term in (\ref{eq:DBITm}) also differs from
\begin{equation}
S_{\mathrm{DBIonic}}=\int\mathrm{d}^4x \sqrt{-f}\left(\Lambda^4\sqrt{1-\Lambda^{-4}(\partial\phi)^2}\right)
\end{equation}
in standard DBIonic screening by an overall sign of the action, which as we will see is also essential for the scalar field to mediate a repulsive fifth force and to drive the late-time acceleration. The second term in the $\mathrm{DBI}T_{\mathrm{m}}$ action (\ref{eq:DBITm}) describes a conformal coupling of the DBI scalar with the trace of the energy-momentum tensor of background matter fields with strength $g\sim\mathcal{O}(1)$ from the stringy perspective.

Suppose the DBI scalar field $\phi(r)$ with a static and spherically symmetric profile is coupled to a static local source $T_{\mathrm{m}}=-\rho_{\mathrm{m}}(r)$; then, the EOM
\begin{equation}
\frac{1}{r^2}\frac{\partial}{\partial r}\left(\frac{r^2\phi'(r)}{\sqrt{1-\Lambda^{-4}\phi'(r)^2}}\right)=-\frac{g}{M_{\mathrm{Pl}}}\rho_{\mathrm{m}}(r)
\end{equation}
can be integrated to give
\begin{equation}
\phi'(r)=-\frac{\Lambda^2}{\sqrt{1+\left(\frac{r}{r_{\mathrm{DBI}}}\right)^4}},
\end{equation}
where a DBI transition radius \cite{Burrage:2014uwa}
\begin{equation}\label{eq:DBIr}
r_{\mathrm{DBI}}=\frac{1}{\Lambda}\left(\frac{gM}{4\pi M_{\mathrm{Pl}}}\right)^{1/2}
\end{equation}
is introduced to separate the DBI regime $r\gg r_{\mathrm{DBI}}$ with repulsive force
\begin{equation}
\vec{a}_{\phi}=-\frac{g}{M_{\mathrm{Pl}}}\phi'(r)\hat{r}\simeq2g^2 G\frac{M}{r^2}\hat{r}=-2g^2\vec{a}_{\mathrm{N}}
\end{equation}
from the Newtonian regime $r\ll r_{\mathrm{DBI}}$ with screened force
\begin{equation}
\vec{a}_{\phi}=-\frac{g}{M_{\mathrm{Pl}}}\phi'(r)\hat{r}\simeq-2g^2\left(\frac{r}{r_{\mathrm{DBI}}}\right)^2\vec{a}_{\mathrm{N}}.
\end{equation}

To retain the success of the MOND phenomenon of DM superfluidity on galactic scales, the DBI force should also be screened from the MOND force on the galactic scale, which renders an identification of the DBI transition radius (\ref{eq:DBIr}) with the MOND transition radius (\ref{eq:MONDr}),
\begin{equation}
r_{\mathrm{DBI}}^2=\frac{1}{\Lambda^2}\frac{gM}{4\pi M_{\mathrm{Pl}}}\Leftrightarrow r_{\mathrm{MOND}}^2=\frac{MG}{a_0}.
\end{equation}
Therefore, the galactic coincidence
\begin{equation}
a_0=\frac{\Lambda^2}{2gM_{\mathrm{Pl}}}\simeq H_0
\end{equation}
is derived, provided that
\begin{equation}
\Lambda^4\simeq M_{\mathrm{Pl}}^2H_0^2\simeq(\mathrm{meV})^4
\end{equation}
for a conformal coupling $g$ of order unity. It turns out as a nice surprise that $\Lambda^4$ coincides with current critical energy density and $\Lambda$ in the $\mathrm{DBI}T_{\mathrm{m}}$ action (\ref{eq:DBITm}) matches that in the $\mathrm{MOND}T_{\mathrm{b}}$ action (\ref{eq:MONDTb}). This is why we use the same symbol for the scale $\Lambda$ in both actions (\ref{eq:MONDTb}) and (\ref{eq:DBITm}), which shares the same scale with the cosmological constant.

\section{DBI dark energy}\label{sec:4}

The repulsive feature of the DBI force and the unexpected match of $\Lambda^4$ with the current critical energy density inspire us to explore the possibility of our DBI scalar field playing the role of dark energy.

We start with the total Lagrangian
\begin{equation}
\sqrt{-f}\mathcal{L}=\sqrt{-f}\mathcal{L}_{\phi}+\sqrt{-f}\mathcal{L}_{\phi T}+\sqrt{-f}\mathcal{L}_{\mathrm{m}},
\end{equation}
where
\begin{align}
\mathcal{L}_{\phi}&=-\Lambda^4\sqrt{1-\Lambda^{-4}(\partial\phi)^2};\\
\mathcal{L}_{\phi T}&=\frac{g\phi}{M_{\mathrm{Pl}}}T_{\mathrm{m}};\\
\mathcal{L}_{\mathrm{m}}&=\mathcal{L}_{\mathrm{m}}(f_{\mu\nu},\psi).
\end{align}

\subsection{Backreaction on matter}

In the absence of the conformal coupling term, the matter component is supposed to behave as a pressureless fluid with the trace $T_{\mathrm{m}}=-\rho_{\mathrm{m}}$ of the energy-momentum tensor $T_{\mu\nu}^{\mathrm{m}}=(2/\sqrt{-f})\delta(\sqrt{-f}\mathcal{L}_{\mathrm{m}})/\delta f^{\mu\nu}$. In the presence of the conformal coupling term, the matter field could exchange momentum by interacting with the DBI scalar field. Therefore, the conformal coupling term would necessarily introduce an effective pressure in the matter fluid, and the effective EOS parameter of matter could in principle deviate from zero. We will show below that such a deviation from pressureless fluid can be made arbitrarily small for a sub-Planckian DBI scalar.

The EOM of the DBI scalar field for a spatial homogenous profile $\phi(t)$ is simply
\begin{equation}\label{eq:EOM}
\ddot{\phi}+3H\dot{\phi}\gamma^{-2}+\frac{gT_{\mathrm{m}}}{M_{\mathrm{Pl}}\gamma^3}=0,
\end{equation}
according to the Euler-Lagrange equation
\begin{equation}
\frac{\partial{(\sqrt{-f}\mathcal{L}_{\phi}+\sqrt{-f}\mathcal{L}_{\phi T})}}{\partial\phi}=\partial_{\mu}\frac{\partial(\sqrt{-f}\mathcal{L}_{\phi}+\sqrt{-f}\mathcal{L}_{\phi T})}{\partial(\partial_{\mu}\phi)}.
\end{equation}

In the absence of the conformal coupling term, the energy-momentum tensor of the DBI scalar field can be computed as
\begin{equation}\label{eq:T1}
T_{\mu\nu}^{\phi}=f_{\mu\nu}\mathcal{L}_{\phi}-\frac{\partial\mathcal{L}_{\phi}}{\partial(\partial^{\mu}\phi)}\partial_{\nu}\phi
\end{equation}
with its energy density and pressure of the form
\begin{align}
\rho_{\phi}=&\Lambda^4\gamma;\\
p_{\phi}=&-\Lambda^4\gamma^{-1}.
\end{align}
In the presence of the conformal coupling term, the conservation equation of the above energy-momentum tensor should be written as
\begin{equation}\label{eq:conservation1}
\nabla^{\mu}T_{\mu\nu}^{\phi}=-\frac{gT_{\mathrm{m}}}{M_{\mathrm{Pl}}}\partial_{\nu}\phi,
\end{equation}
where the temporal component of the above equation reads
\begin{equation}\label{eq:conservation10}
\dot{\rho}_{\phi}+3H(\rho_{\phi}+p_{\phi})=-\frac{g\rho_{\mathrm{m}}}{M_{\mathrm{Pl}}}\dot{\phi},
\end{equation}
which is consistent with the EOM (\ref{eq:EOM}).

In the absence of the conformal coupling term, the EOM (\ref{eq:EOM}) has a trivial solution $\dot{\phi}=0$, and the EOS parameter
\begin{equation}
w_{\phi}=\frac{p_{\phi}}{\rho_{\phi}}=-\gamma^{-2}\equiv-1-\Lambda^{-4}\dot{\phi}^2
\end{equation}
would simply imply a cosmological constant with $w_{\phi}=-1$. In the presence of the conformal coupling term, the EOM (\ref{eq:EOM}) cannot admit such a trivial solution $\dot{\phi}=0$ unless $\phi$ is always equal to zero, which is of less physical interest. Therefore, our DBI scalar should generally behave as a dynamical Chaplygin gas \cite{Kamenshchik:2001cp} $p_{\phi}=-\Lambda^8/\rho_{\phi}$ with phantomlike EOS parameter and superluminal sound speed \cite{Mukhanov:2005bu} $c_s^2=\dot{p}/\dot{\rho}=\gamma^{-2}$, where the closed timelike curves are argued to be evaded within the regime of validity of the effective field theory (EFT) due to chronology protection \cite{Burrage:2011cr,Babichev:2007dw}. With slow-roll condition $\dot{\phi}\ll\Lambda^2$, our DBI scalar could serve as a candidate for the DE sector. We will show below that such a slow-roll condition can be satisfied for a sub-Planckian DBI scalar as well.

To derive the conservation equation for the matter component, we start with an alternative definition of the energy-momentum tensor for the DBI scalar,
\begin{equation}\label{eq:T2}
T_{\mu\nu}^{\phi+\phi T}=f_{\mu\nu}(\mathcal{L}_{\phi}+\mathcal{L}_{\phi T})-\frac{\partial(\mathcal{L}_{\phi}+\mathcal{L}_{\phi T})}{\partial(\partial^{\mu}\phi)}\partial_{\nu}\phi,
\end{equation}
with its energy density and pressure of the form
\begin{align}
\rho_{\phi T}=&\Lambda^4\gamma+\frac{g\phi}{M_{\mathrm{Pl}}}\rho_{\mathrm{m}};\\
p_{\phi T}=&-\Lambda^4\gamma^{-1}-\frac{g\phi}{M_{\mathrm{Pl}}}\rho_{\mathrm{m}}.
\end{align}
In the presence of the conformal coupling term, the conservation equation of the above energy-momentum tensor should be written as
\begin{equation}\label{eq:conservation2}
\nabla^{\mu}T_{\mu\nu}^{\phi+\phi T}=\frac{g\phi}{M_{\mathrm{Pl}}}\partial_{\nu}T_{\mathrm{m}},
\end{equation}
where the temporal component of the above equation reads
\begin{equation}\label{eq:conservation20}
\dot{\rho}_{\phi T}+3H(\rho_{\phi T}+p_{\phi T})=\frac{g\phi}{M_{\mathrm{Pl}}}\dot{\rho}_{\mathrm{m}},
\end{equation}
which is also consistent with the EOM (\ref{eq:EOM}).

Since the total energy-momentum tensor is conserved, the conservation equation of the energy-momentum tensor of the matter component is thus
\begin{equation}\label{eq:conservation3}
\nabla^{\mu}T_{\mu\nu}^{\mathrm{m}}=-\frac{g\phi}{M_{\mathrm{Pl}}}\partial_{\nu}T_{\mathrm{m}},
\end{equation}
where the temporal component of the above equation reads
\begin{equation}\label{eq:conservation30}
\dot{\rho}_{\mathrm{m}}+3H\rho_{\mathrm{m}}=-\frac{g\phi}{M_{\mathrm{Pl}}}\dot{\rho}_{\mathrm{m}}.
\end{equation}
The source term on the right-hand side of above equation can be accounted for by recognizing the effective EOS parameter of the matter component as
\begin{equation}
w_{\mathrm{m}}=\frac{1}{1+\frac{g\phi}{M_{\mathrm{Pl}}}}-1.
\end{equation}
Therefore, the backreaction of the DBI field on the matter component due to the conformal coupling term can be safely neglected in the field region $\phi\ll M_{\mathrm{Pl}}$ of the DBI scalar for conformal coupling of order unity. From now on, we will take a fiducial value $g=1$ for the conformal coupling in order to solve the galactic coincidence problem.

\subsection{Steady flow assumption}

In the rest of this section, we will work with the assumption, called the \emph{steady flow} assumption, that the energy flow from the DBI scalar to the matter component is conserved. We define the energy flow as the energy-momentum tensor associated with the conformal coupling term
\begin{equation}
T_{\mu\nu}^{\phi T}=T_{\mu\nu}^{\phi+\phi T}-T_{\mu\nu}^{\phi}=f_{\mu\nu}\mathcal{L}_{\phi T};
\end{equation}
then, steady flow assumption is expressed as
\begin{equation}\label{eq:conservation4}
\nabla^{\mu}T_{\mu\nu}^{\phi T}=\frac{g}{M_{\mathrm{Pl}}}\partial_{\nu}(\phi T_{\mathrm{m}})=0,
\end{equation}
where the temporal component of the above equation reads
\begin{equation}\label{eq:conservation40}
\dot{\phi}\rho_{\mathrm{m}}+\phi\dot{\rho}_{\mathrm{m}}=0.
\end{equation}
The steady flow assumption simply states that, although the energy-momentum tensors of the DBI field and matter field are not separately conserved as indicated in Eqs. (\ref{eq:conservation1}) and (\ref{eq:conservation3}), there is no loss during the energy transfer from the DBI scalar to the matter component and the total energy-momentum tensor of the DBI field and the matter field is conserved, namely, $\nabla^{\mu}T_{\mu\nu}^{\phi}+\nabla^{\mu}T_{\mu\nu}^{\mathrm{m}}=-\nabla^{\mu}T_{\mu\nu}^{\phi T}=0$. We will justify numerically the steady flow assumption below.

\begin{figure*}
  \includegraphics[width=8cm]{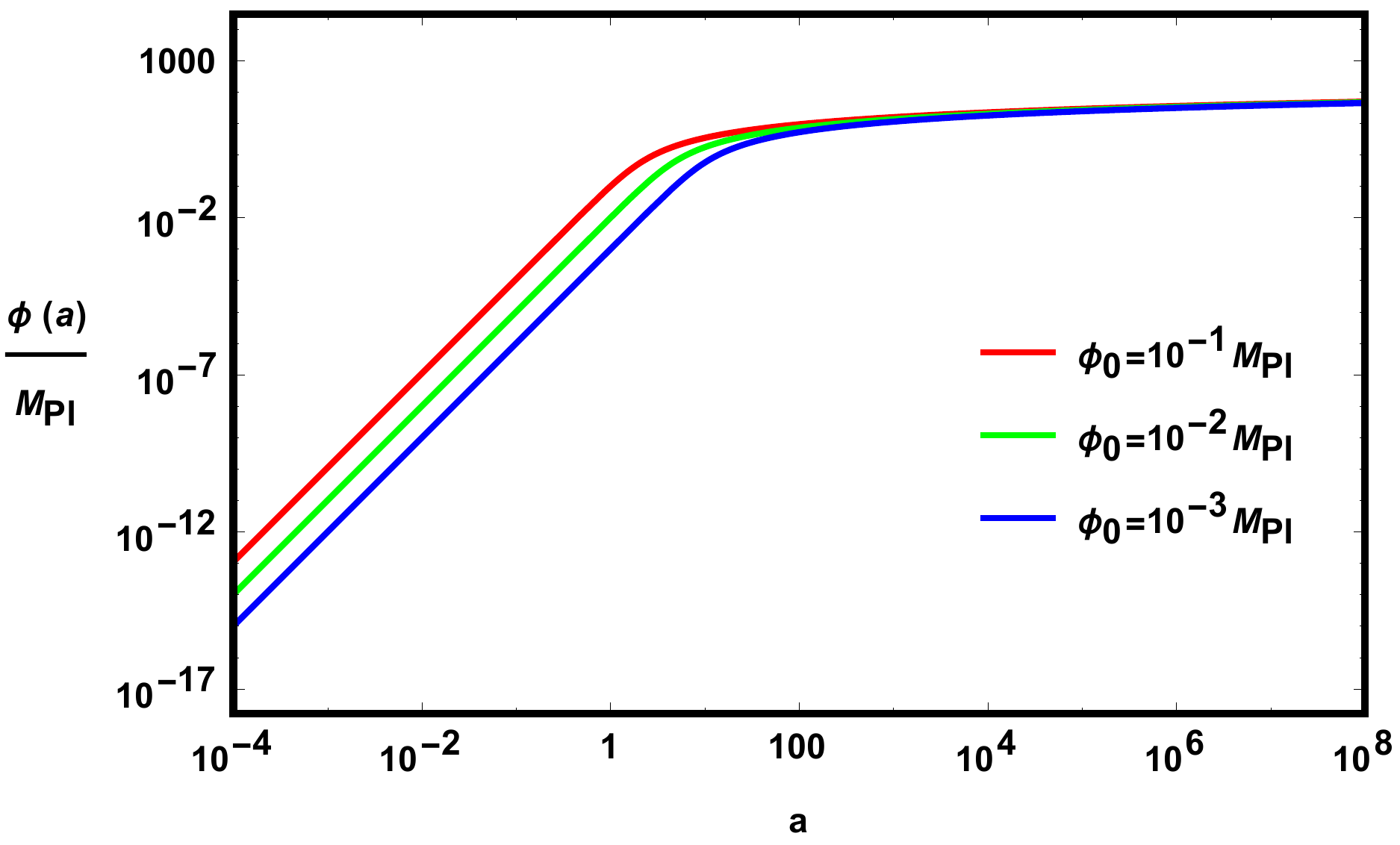}
  \includegraphics[width=8cm]{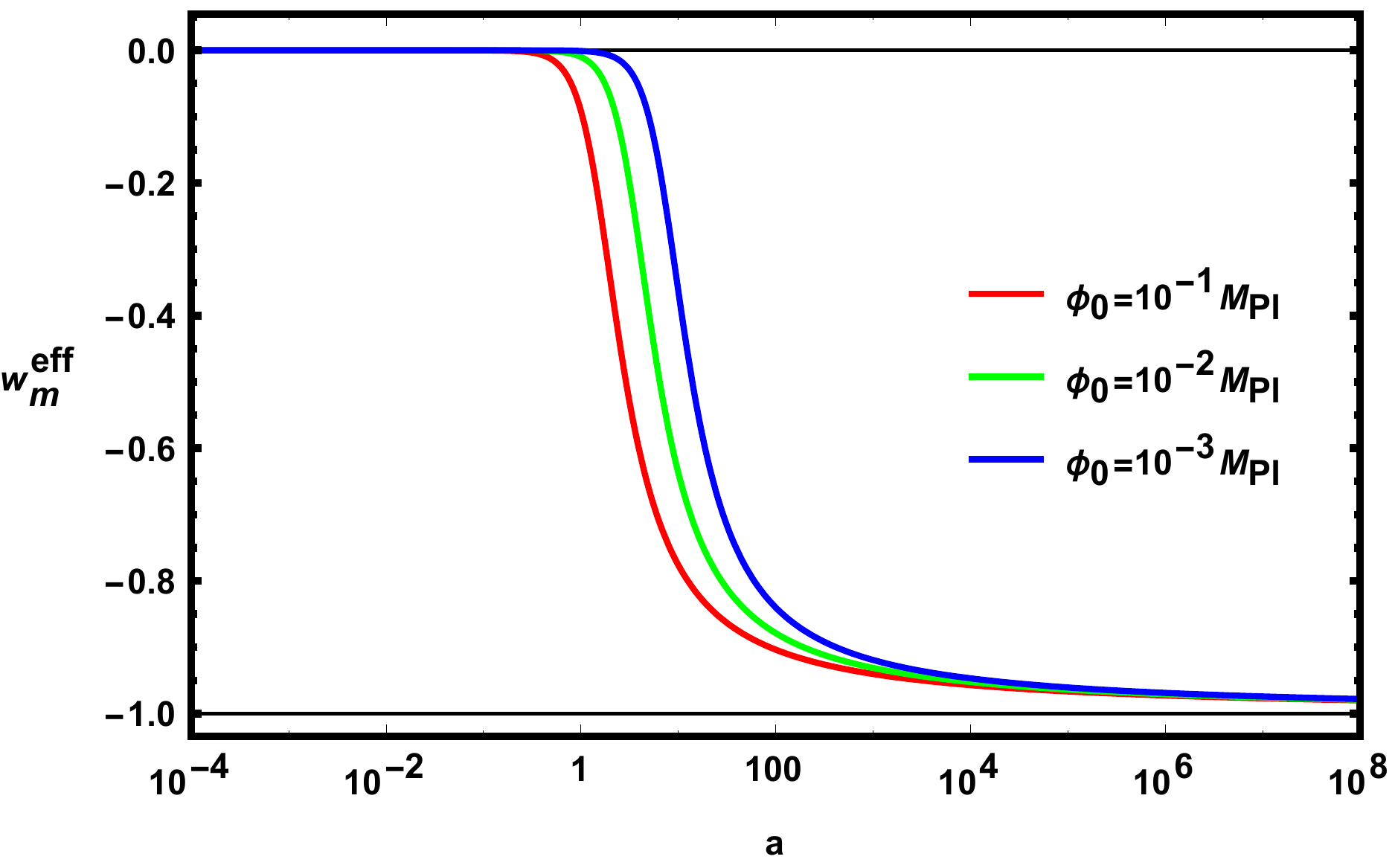}\\
  \includegraphics[width=8cm]{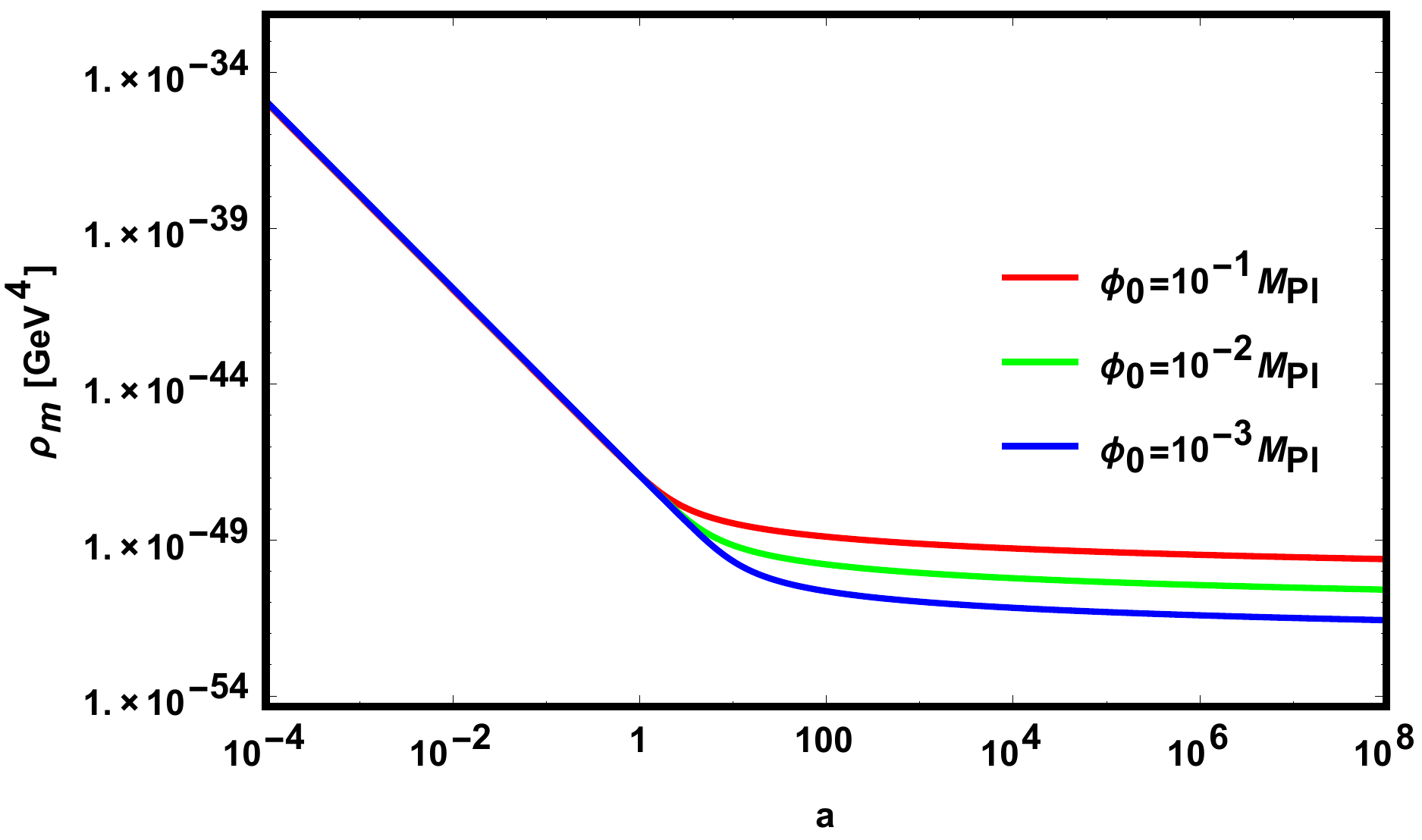}
  \includegraphics[width=8cm]{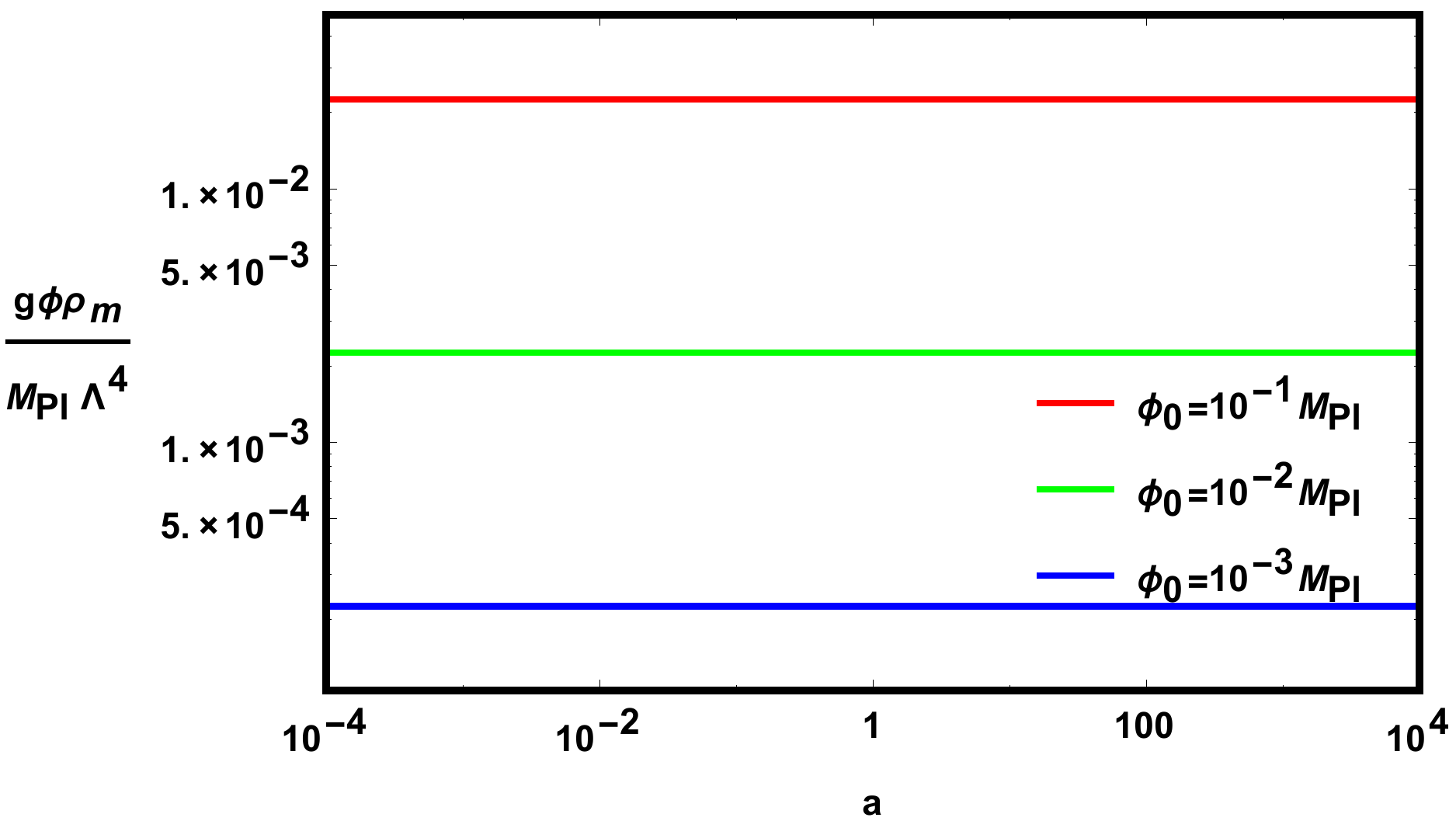}\\
  \caption{The evolutions of the DBI field, the effective EOS parameter of the matter component, the matter energy density, and the conformal coupling term with respect to the scale factor for initial conditions $\phi_0/M_{\mathrm{Pl}}=10^{-1},10^{-2},10^{-3}.$}\label{fig:phi and so on}
\end{figure*}

With the steady flow assumption, one can solve the DBI field
\begin{equation}
\phi(a)=\frac{M_{\mathrm{Pl}}}{g}W\left(\frac{g\phi_0}{M_{\mathrm{Pl}}}e^{\frac{g\phi_0}{M_{\mathrm{Pl}}}}\left(\frac{a}{a_0}\right)^3\right)
\end{equation}
analytically by combining Eq. (\ref{eq:conservation30}) with Eq. (\ref{eq:conservation40}), where $\phi_0\equiv\phi(a=a_0)$ with present-day scale factor $a_0\equiv1$ and $W(z)$ is the Lambert W function defined by $z=W(z)\exp[W(z)]$.  Hence, the evolution equation (\ref{eq:conservation30}) of the matter component can be directly integrated to give
\begin{equation}
\rho_{\mathrm{m}}(a)=\rho_{\mathrm{m}0}\exp\left(-3\int_{a_0}^{a}\frac{\mathrm{d}\ln a'}{1+W\left(\frac{g\phi_0}{M_{\mathrm{Pl}}}e^{\frac{g\phi_0}{M_{\mathrm{Pl}}}}\left(\frac{a'}{a_0}\right)^3\right)}\right).
\end{equation}
The evolutions of DBI field, the effective EOS parameter of matter component, the matter energy density, and the conformal coupling term are presented in Fig. \ref{fig:phi and so on}
The backreaction of the DBI field on the matter component is negligible during the matter dominated era as long as a sub-Planckian field value for the DBI field at present is specified. However, the effective EOS parameter of the matter component will eventually approach $-1$ in the future, causing an unavoidable vacuum decay to matter, saving us from big rip singularity as we will see. The steady flow assumption is justified by a constant conformal coupling term. At small scale factor $a\ll1$, the evolution of the Lambert W function $W(a^3)\sim a^3$ compensates the evolution of the matter component $\rho_{\mathrm{m}}\sim a^{-3}$ to render a constant conformal coupling term $\phi T_{\mathrm{m}}\sim W(a^3)\rho_{\mathrm{m}}\sim\mathrm{const}$. At a large scale factor, the constant nature of the conformal coupling term is nontrivial.

\begin{figure*}
  \includegraphics[width=8cm]{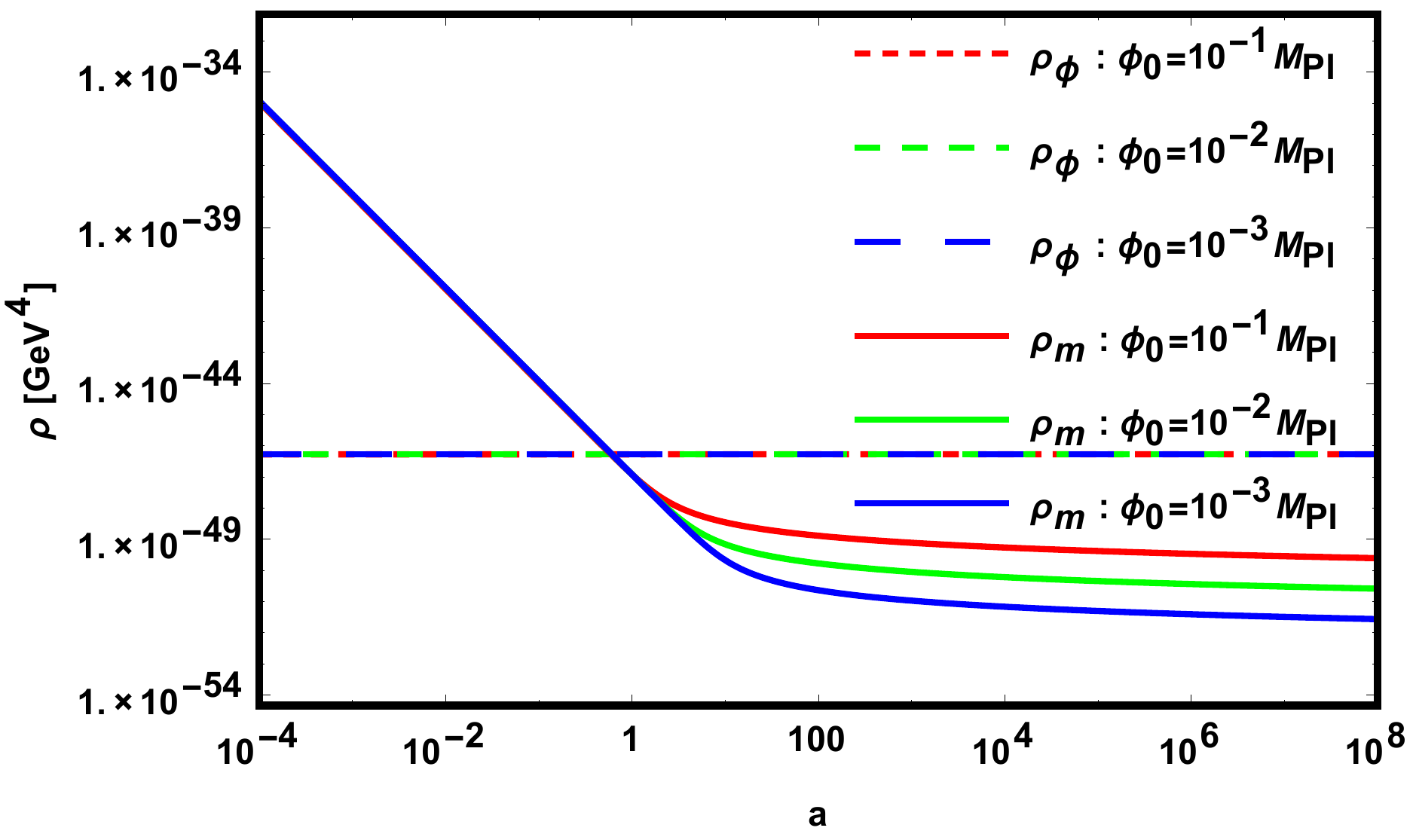}
  \includegraphics[width=8cm]{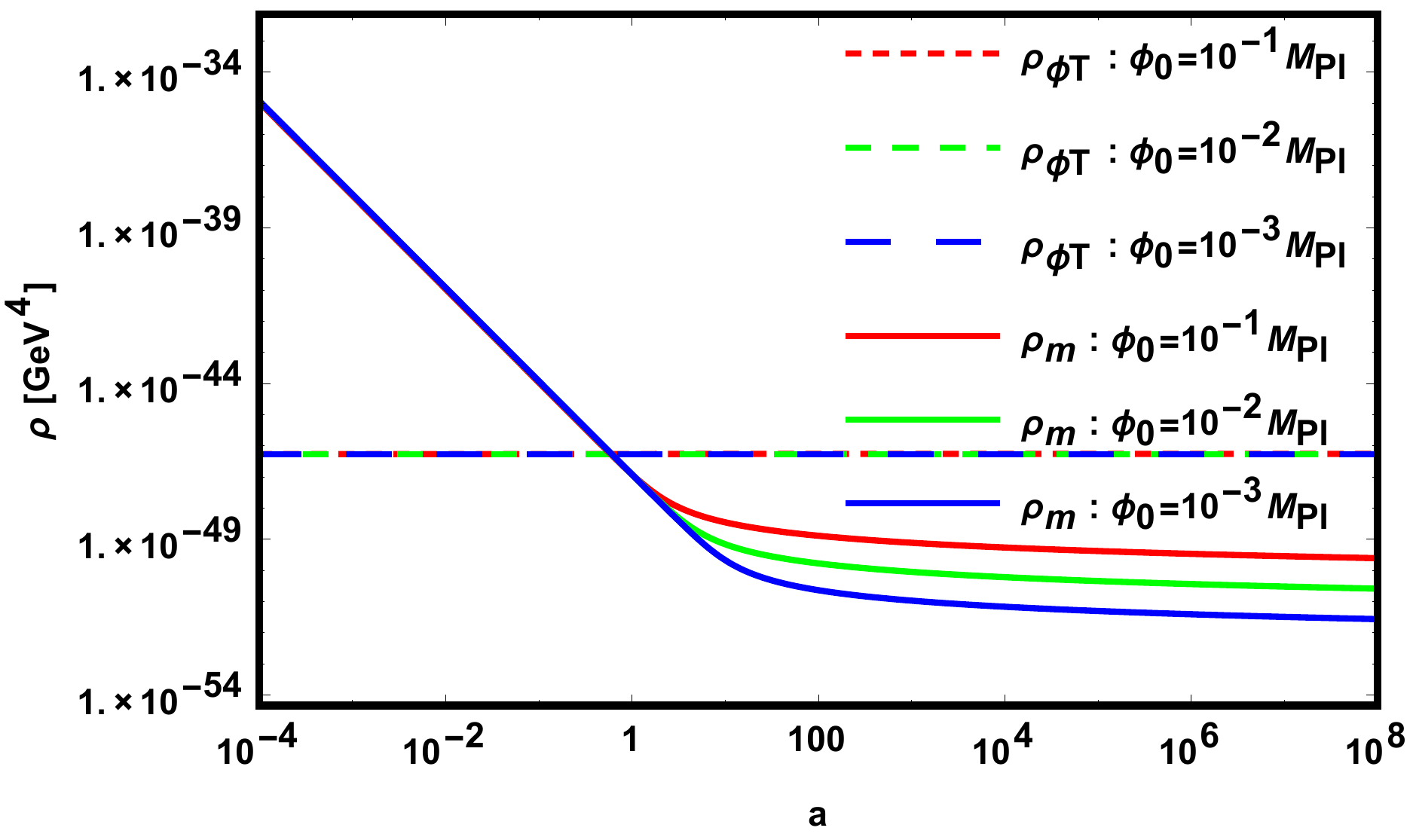}\\
  \includegraphics[width=8cm]{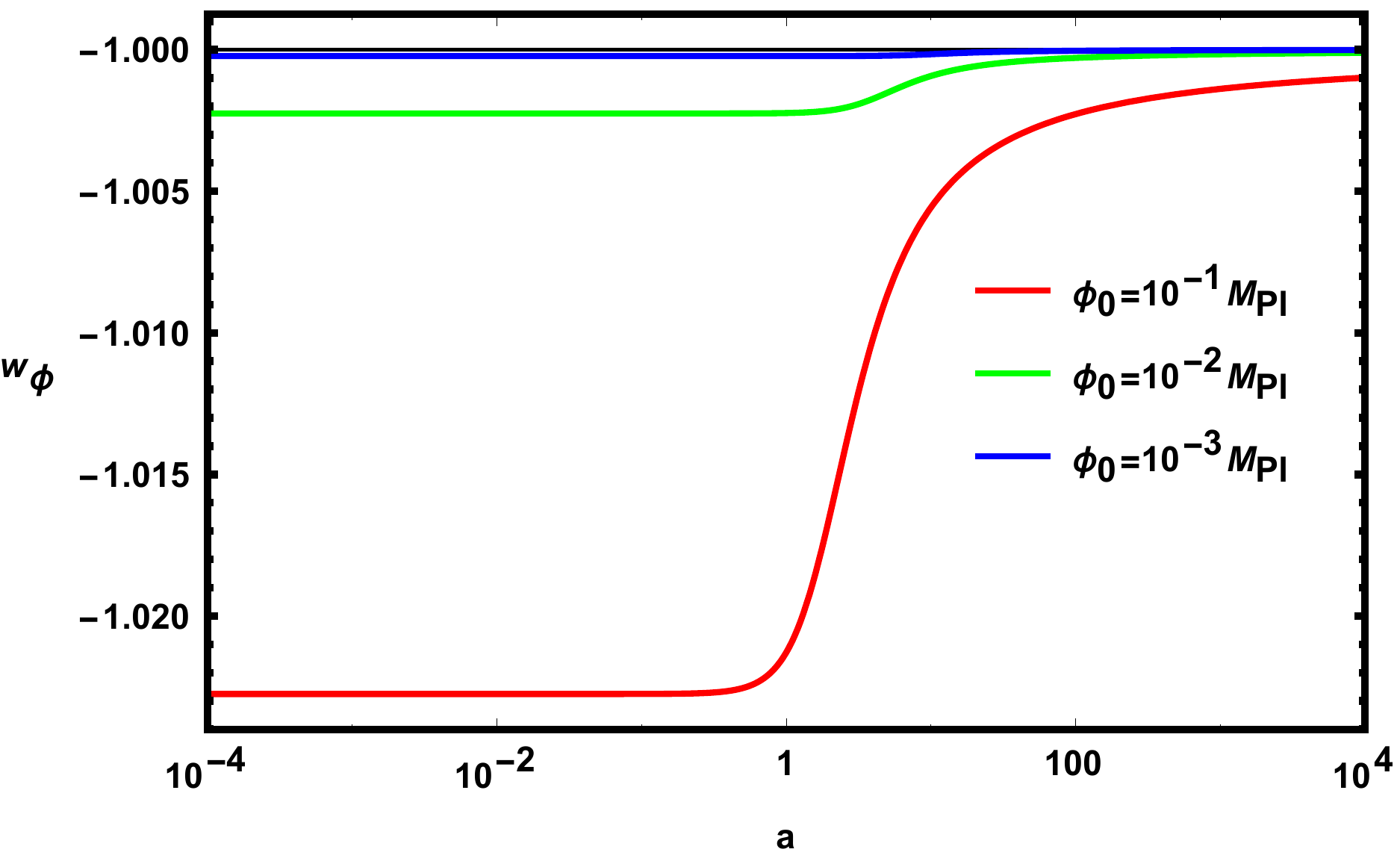}
  \includegraphics[width=8cm]{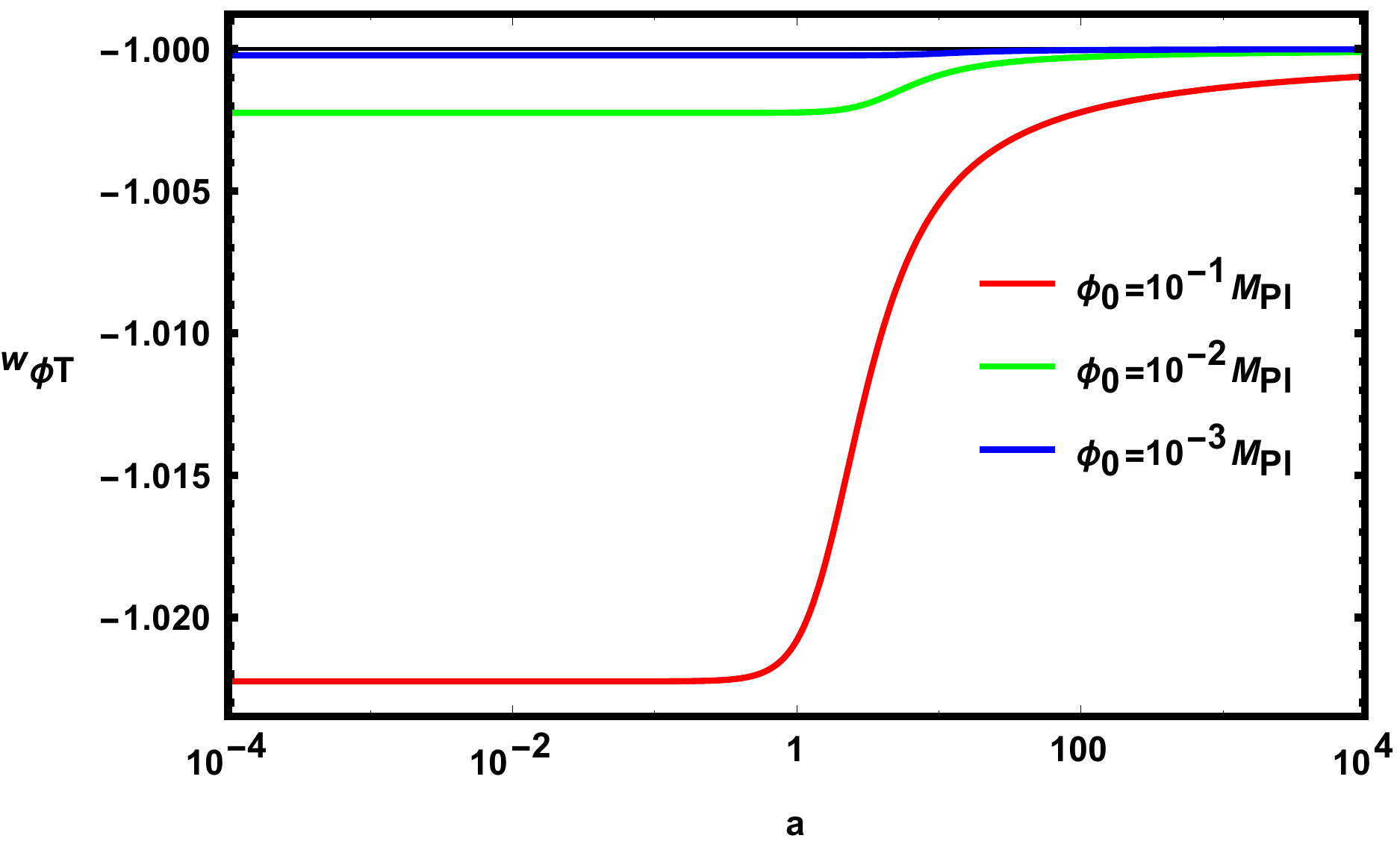}\\
  \includegraphics[width=8cm]{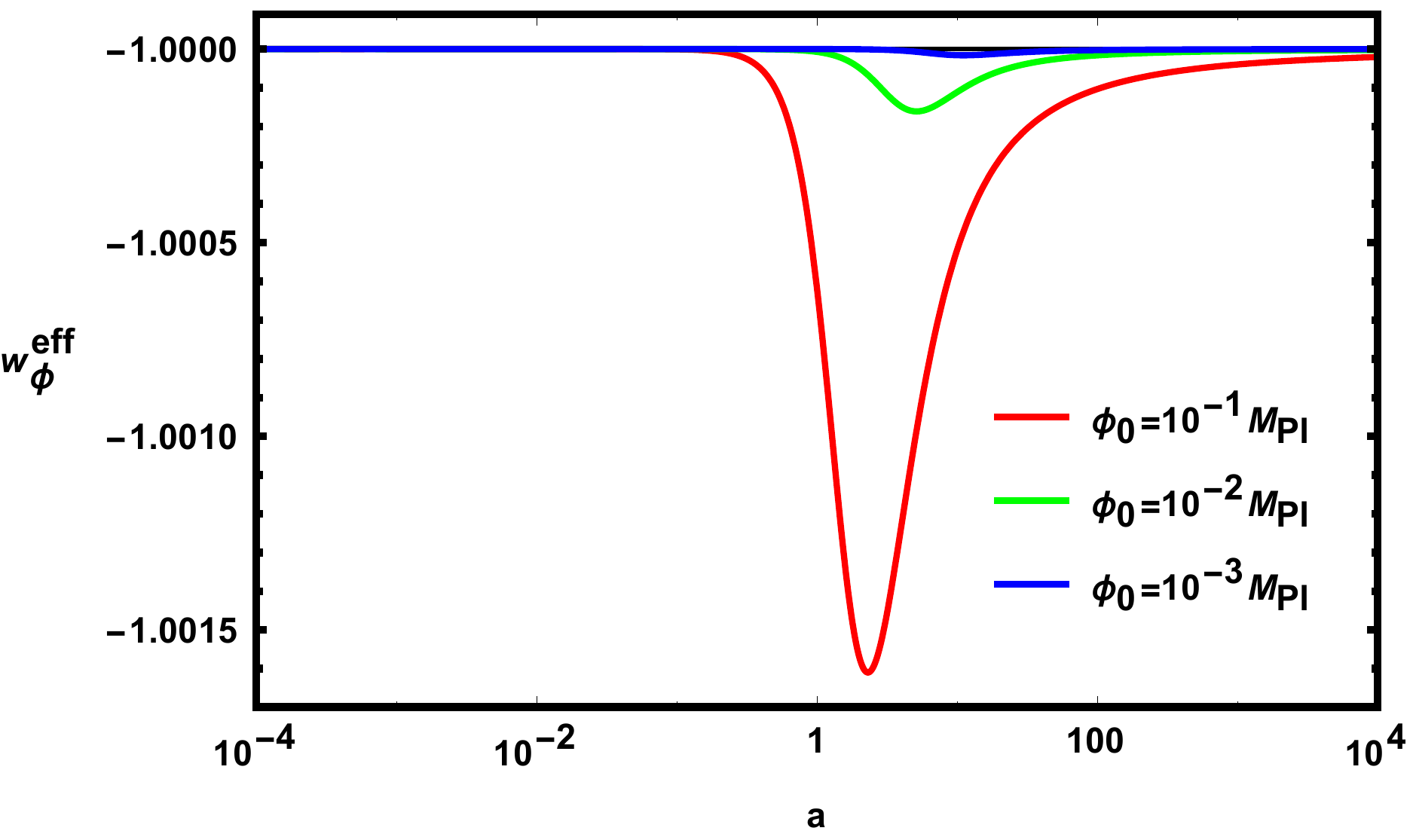}
  \includegraphics[width=8cm]{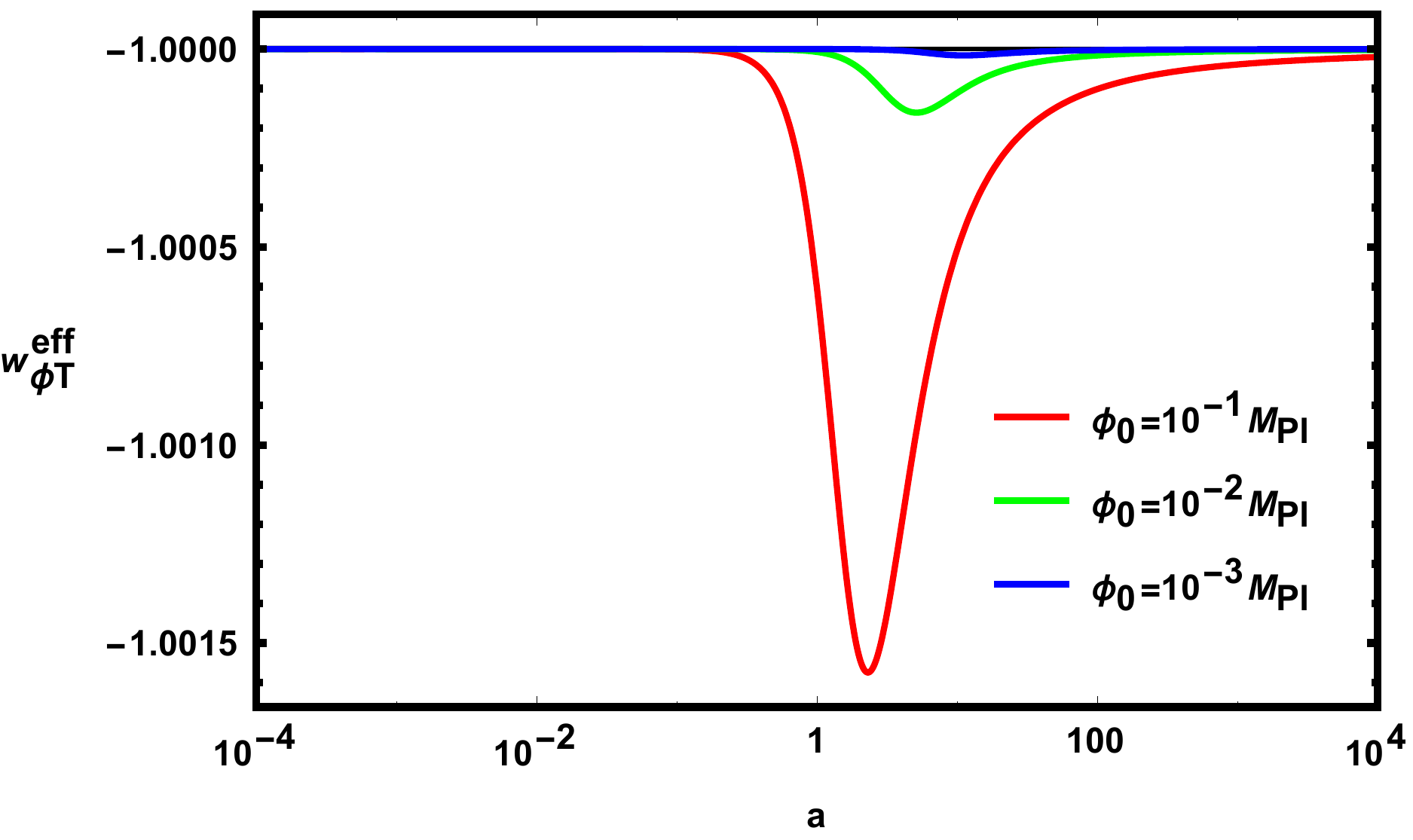}\\
  \caption{The evolutions of the energy density of the DBI field and their effective EOS parameters with respect to the scale factor for initial conditions $\phi_0/M_{\mathrm{Pl}}=10^{-1},10^{-2},10^{-3}.$}\label{fig:rhophi and so on}
\end{figure*}

\begin{figure*}
  \includegraphics[width=8cm]{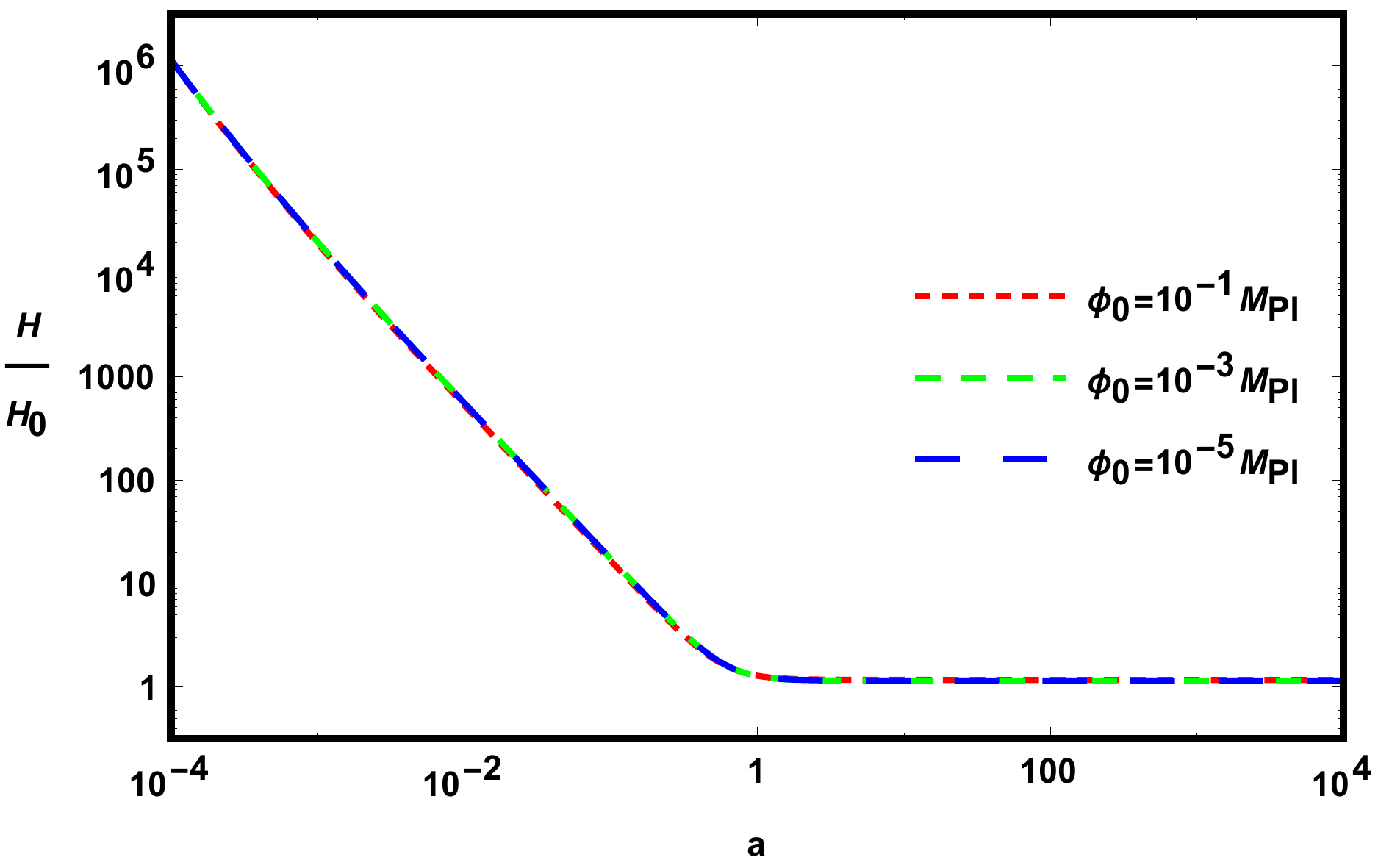}
  \includegraphics[width=8cm]{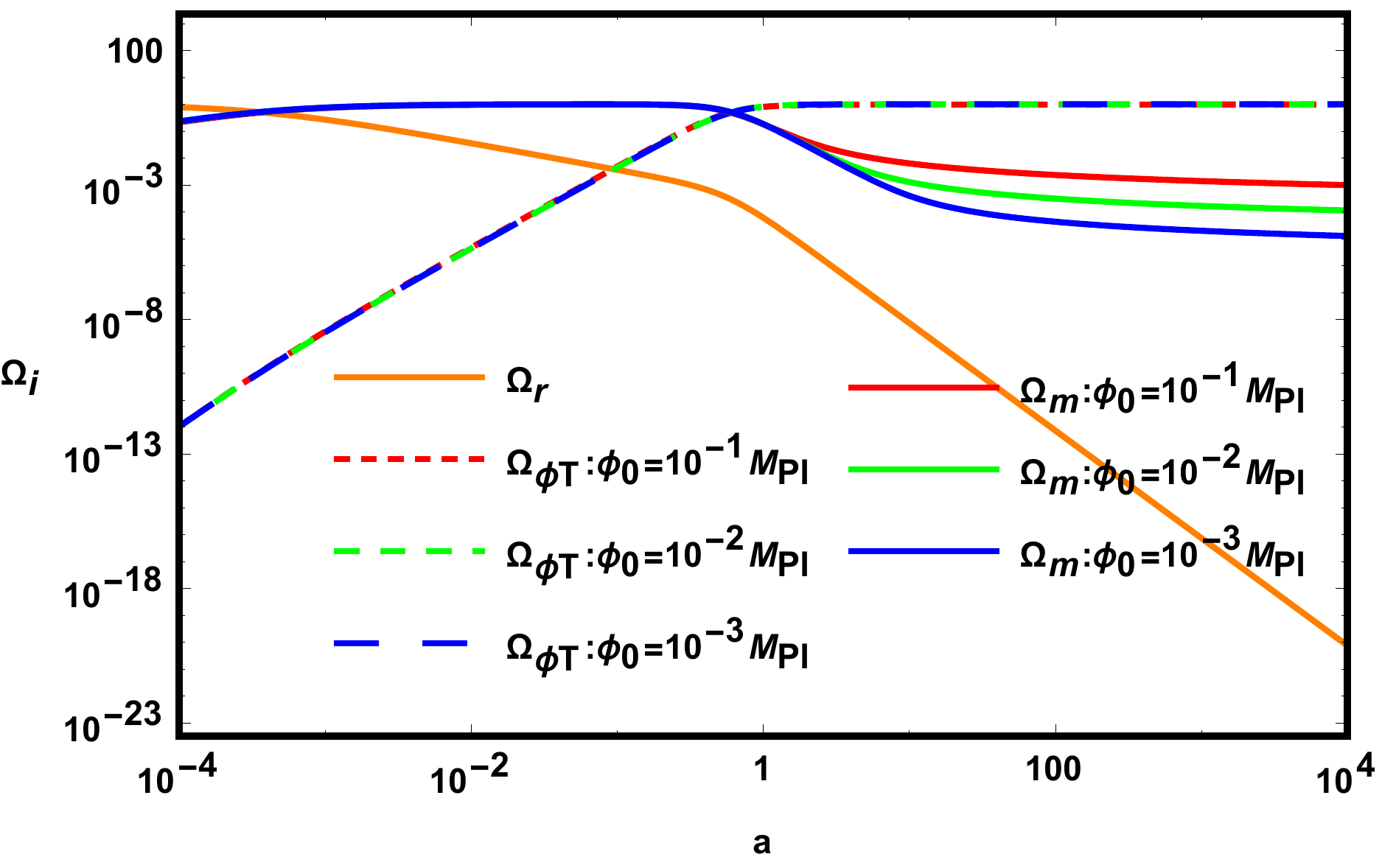}\\
  \caption{The evolutions of the Hubble parameter and fractions of energy density with respect to the scale factor for initial conditions $\phi_0/M_{\mathrm{Pl}}=10^{-1},10^{-2},10^{-3}.$}\label{fig:Hubble and Omega}
\end{figure*}

The evolution of the energy density of the DBI field can be solved numerically by rewriting Eq. (\ref{eq:conservation10}) as
\begin{equation}
\rho'_{\phi}(a)+\frac{3}{a}\left(\rho_{\phi}(a)-\frac{\Lambda^8}{\rho_{\phi}(a)}\right)=-\frac{g\rho_{\mathrm{m}}(a)}{M_{\mathrm{Pl}}}\phi'(a).
\end{equation}
With numerical solution $\rho_{\phi}(a)$, one can evaluate all other quantities like
\begin{align}
w_{\phi}(a)&=-\left(\Lambda^{-4}\rho_{\phi}(a)\right)^{-2};\\
w_{\phi}^{\mathrm{eff}}(a)&=w_{\phi}(a)+\frac{g a}{3M_{\mathrm{Pl}}}\phi'(a)\frac{\rho_{\mathrm{m}}(a)}{\rho_{\phi}(a)};\\
\rho_{\phi T}(a)&=\rho_{\phi}(a)+\frac{g}{M_{\mathrm{Pl}}}\phi(a)\rho_{\mathrm{m}}(a);\\
w_{\phi T}(a)&=\frac{-\frac{\Lambda^8}{\rho_{\phi}(a)}-\frac{g}{M_{\mathrm{Pl}}}\phi(a)\rho_{\mathrm{m}}(a)}{\rho_{\phi}(a)+\frac{g}{M_{\mathrm{Pl}}}\phi(a)\rho_{\mathrm{m}}(a)};\\
w_{\phi T}^{\mathrm{eff}}(a)&=w_{\phi T}(a)-\frac{g a}{3M_{\mathrm{Pl}}}\phi(a)\frac{\rho'_{\mathrm{m}}(a)}{\rho_{\phi T}(a)},
\end{align}
where the effective EOS parameters $w_{\phi}^{\mathrm{eff}}(a)$ and $w_{\phi T}^{\mathrm{eff}}(a)$ of the DBI scalar field are defined by rewriting Eqs. (\ref{eq:conservation10}) and (\ref{eq:conservation20}) in a form without the interacting term,
\begin{align}
&\dot{\rho}_{\phi}+3H(1+w_{\phi}^{\mathrm{eff}})\rho_{\phi}=0;\\
&\dot{\rho}_{\phi T}+3H(1+w_{\phi T}^{\mathrm{eff}})\rho_{\phi T}=0.
\end{align}
The evolutions of the above quantities are plotted in Fig. \ref{fig:rhophi and so on}.
The division of DBI fluid from matter fluid is somewhat artificial since the DBI scalar and matter component are coupled together. However, the difference between definitions (\ref{eq:T1}) and (\ref{eq:T2}) of the energy-momentum tensor of the DBI scalar are shown to be negligible in Fig. \ref{fig:rhophi and so on}; therefore, we will just stick to Eq. (\ref{eq:T1}) for the sake of simplicity. We also compute the evolution of the Hubble parameter by $3M_{\mathrm{Pl}}^2H(a)^2=\rho_{\phi T}(a)+\rho_{\mathrm{m}}(a)+\rho_{\mathrm{r}}(a)$ and the fractions of energy density by $\Omega_i(a)=\rho_i(a)/3M_{\mathrm{Pl}}^2H(a)^2$ in Fig. \ref{fig:Hubble and Omega}.
It is worth noting that the DBI scalar relaxes its phantom nature by vacuum decaying to matter, preventing the matter component from being diluted away and leading to a constant Hubble parameter in the asymptotic future free of big rip singularity.

\subsection{Slow-roll conditions}

\begin{figure}
  \includegraphics[width=8cm]{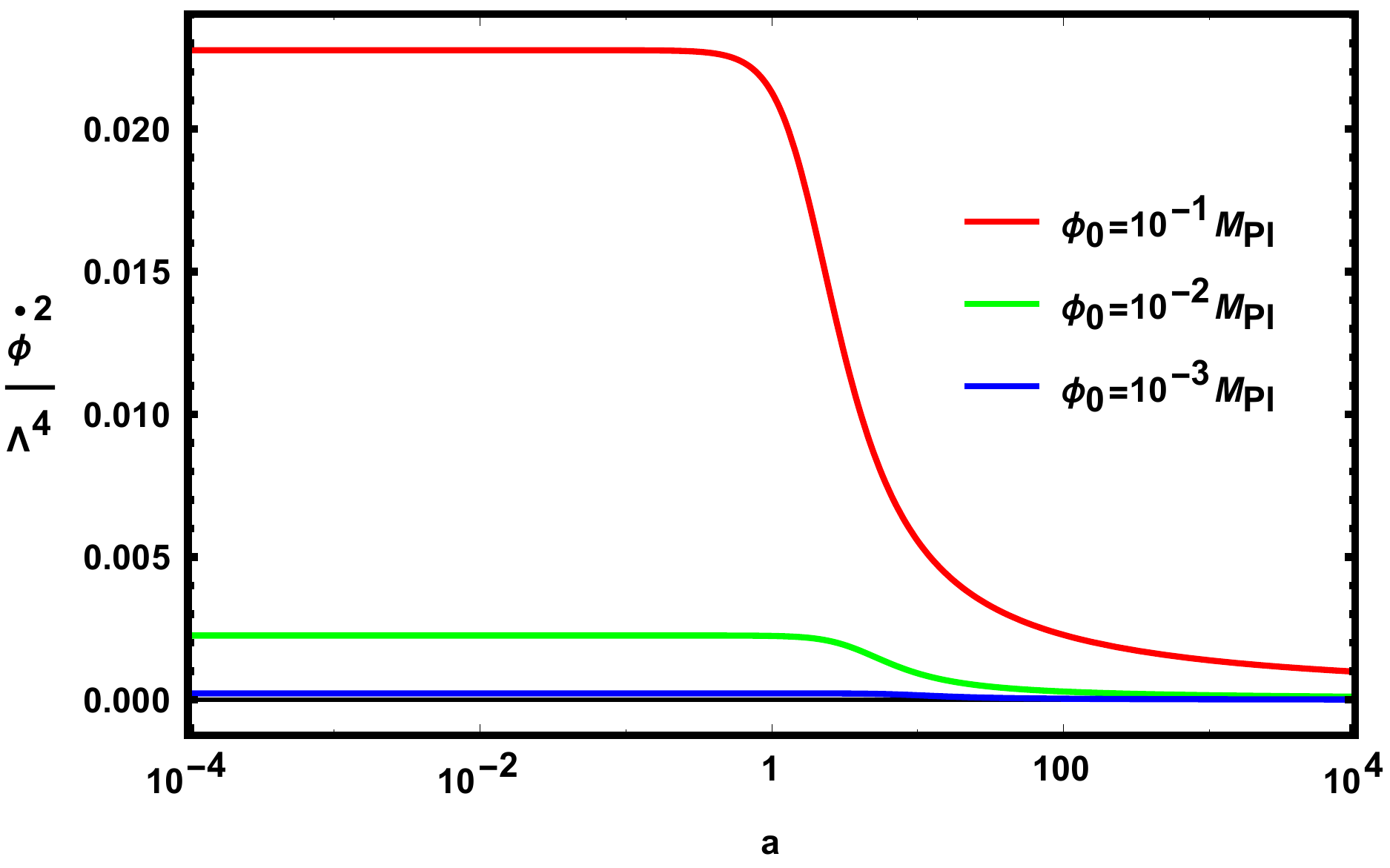}\\
  \includegraphics[width=8cm]{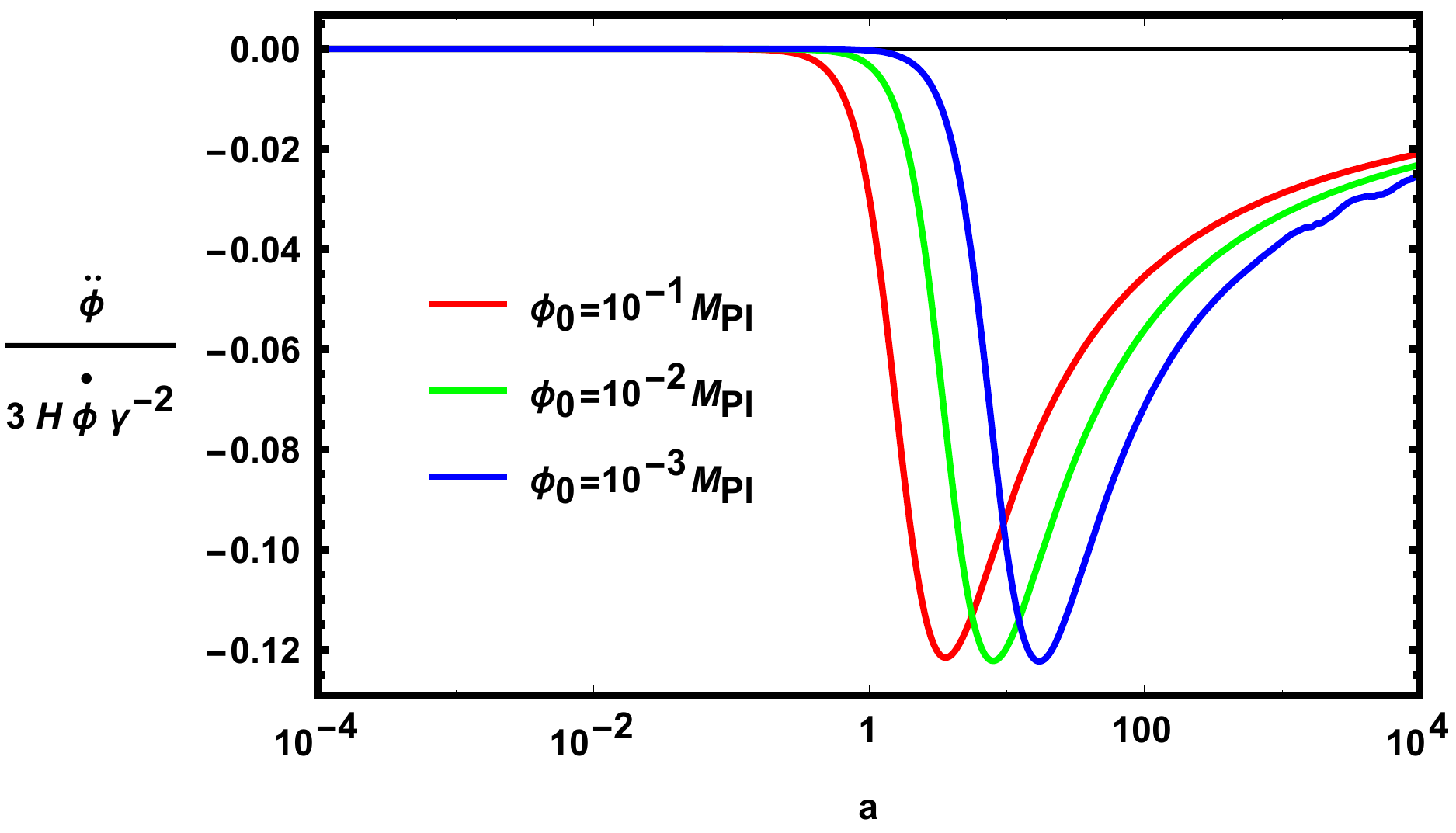}\\
  \includegraphics[width=8cm]{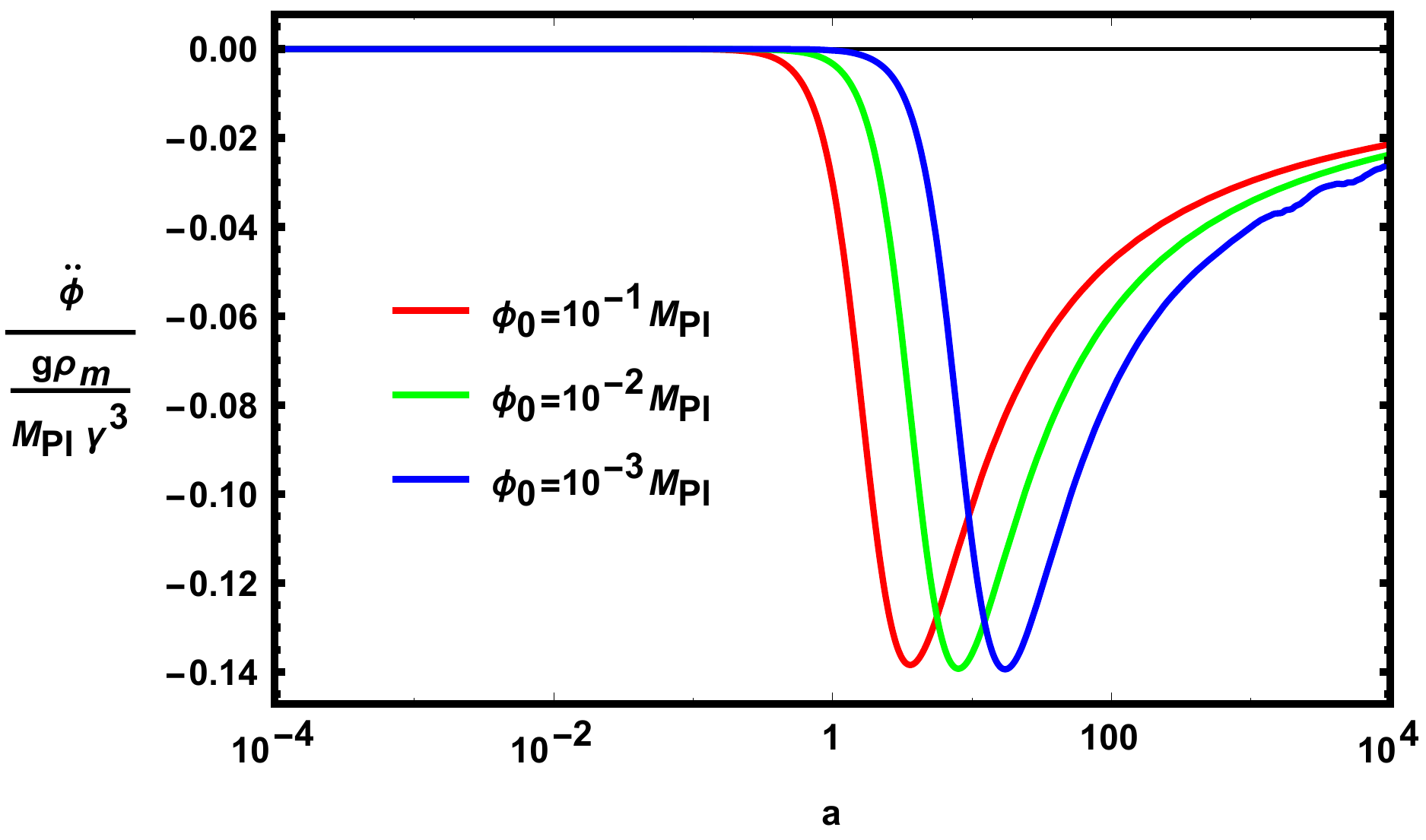}\\
  \caption{The evolutions of the first slow-roll condition $\dot{\phi}^2\ll\Lambda^4$ and the second slow-roll condition $|\ddot{\phi}|\ll3H\dot{\phi}\gamma^{-2}, |\ddot{\phi}|\ll\frac{g\rho_{\mathrm{m}}}{M_{\mathrm{Pl}}\gamma^3}$, with respect to the scale factor for initial conditions $\phi_0/M_{\mathrm{Pl}}=10^{-1},10^{-2},10^{-3}.$}\label{fig:slow-roll}
\end{figure}

Last but not least, it is the slow-roll condition
\begin{equation}\label{eq:slow-roll 1}
\frac{\dot{\phi}^2}{\Lambda^4}\ll1
\end{equation}
that allows us to interpret our DBI scalar as a candidate for the dark energy sector.
To evaluate analytically the EOS parameter of our DBI DE, we propose a second slow-roll condition,
\begin{equation}\label{eq:slow-roll 2}
\left|\frac{\ddot{\phi}}{3H\dot{\phi}\gamma^{-2}}\right|\ll1,\left|\frac{\ddot{\phi}}{\frac{g\rho_{\mathrm{m}}}{M_{\mathrm{Pl}}\gamma^3}}\right|\ll1,
\end{equation}
on the EOM (\ref{eq:EOM}) and find that
\begin{equation}
\dot{\phi}^2\simeq\frac{g^2T_{\mathrm{m}}^2}{9M_{\mathrm{Pl}}^2H^2\gamma^2}.
\end{equation}
Recalling that the factor $\gamma\equiv1/\sqrt{1+\Lambda^{-4}\dot{\phi}^2}$ and the matter component $T_{\mathrm{m}}=-\rho_{\mathrm{m}}=-3M_{\mathrm{Pl}}^2H^2\Omega_{\mathrm{m}}$ and the galactic coincidence $\Lambda^4=4g^2M_{\mathrm{Pl}}^2H_0^2$, one can immediately derive from the above equation the EOS parameter
\begin{equation}\label{eq:EOS}
w_{\phi}=-\gamma^{-2}\simeq\frac{1}{-1+E^2\Omega_{\mathrm{m}}^2/4},
\end{equation}
where the reduced Hubble parameter $E=H/H_0$ is understood and the conformal coupling $g$ is surprisingly canceled out. Testing Eq. (\ref{eq:EOS}) with the present value of matter fraction $\Omega_{\mathrm{m}0}\approx0.3$, one finds the present value of the EOS of our DBI DE,
\begin{equation}\label{eq:w0}
w_{\phi0}\simeq\frac{1}{-1+\Omega_{\mathrm{m}0}^2/4}\approx-1.023,
\end{equation}
perfectly matching the Planck 2015 constraints \cite{Ade:2015xua}. A distinct feature of our DBI DE is that $w_{\phi0}$ and $\Omega_{\mathrm{m}0}$ are strongly correlated without other free parameters encountered. Although behaving mildly like the phantom at present, our DBI DE will relax its phantom nature by vacuum decaying to matter, preventing matter from being diluted away, resulting in a constant Hubble parameter and leading to a de Sitter future free of big rip singularity. The validity of the first and second slow-roll conditions (\ref{eq:slow-roll 1}) and (\ref{eq:slow-roll 2}) is presented in Fig. \ref{fig:slow-roll}.

\section{Conclusions and discussions}\label{sec:5}

It was recently claimed that the axionlike dark matter particles can condense on galactic scales as a superfluid, the phonons of which mediate MONDian force between baryons, and thus MOND arises as an emergent phenomenon of dark matter itself. The standard $\Lambda$CDM model is recovered on cosmic scales in the presence of dark matter particles in the normal phase instead of the condensed phase. We have proposed to study the possible origin of the MOND critical acceleration scale in the context of dark matter superfluidity. We have introduced a DBI-like scalar field conformally coupled to the matter components. It turns out that the MOND critical acceleration is roughly at the same magnitude with the present Hubble scale, provided that the conformally coupled DBI scalar plays the role of dark energy.

However, one might be concerned with the possible ghost problem of our proposal. In canonical quantum field theory, a Lagrangian with a wrong-sign kinetic term, after canonical quantization, usually admits the negative norm states with negative energy, namely, the ghost states. If there are no other fields directly coupled to the ghost field, it would not cause us any trouble. However, if there are other fields with a correct-sign kinetic term directly coupled to the ghost field, the vacuum would be unstable because it could generate a pair of ghost particles with negative energy and a pair of normal particles with positive energy. We argue that the possible ghost problem might not be as pronounced as it appears to be due to the following three features encountered in our model. First, the Hamiltonian density turns out to be positive and bounded below, which suggests that there might be a stable vacuum where ghost particles can condense. Second, the equation of motion is second order in the time derivative, which might evade the ghost problem from the view point of Ostrogradsky's theorem. Third, even if the ghosts indeed exist, they are indirectly coupled to the matter fields via the trace of the energy-momentum tensor. Since the matter fields act as a source term, there are simply no sources for ghosts to be generated when DBI-like scalar field come to dominate. This might explain why the equation of state of our DBI dark energy approaches $-1$ in the end. Therefore, our model should be treated as a phenomenological model which requires further study in the future.

\begin{acknowledgments}
S.J.W. would like to thank Lasha Berezhiani and Alexander Vikman for helpful correspondences and Bin Hu, Jian-Wei Hu, Qi Guo, and Run-Qiu Yang for helpful discussions. We would like to thank an anonymous referee for greatly improving the presentation and validity of the paper.
R.G.C. is supported by the Strategic Priority Research Program of the Chinese Academy of Sciences, Grant No.XDB09000000.
\end{acknowledgments}

\bibliographystyle{apsrev4-1}
\bibliography{ref}

\end{document}